# Specifics of thermodynamic description of nanocrystals

A I Karasevskii

G.V. Kurdyumov Institute for Metal Physics of Ukrainian Academy of Sciences

36 Vernadsky bl., Kiev, 03142, Ukraine

E-mail: akaras@imp.kiev.ua

**Abstract**

A method of statistical description of thermodynamic properties of nanocrystals is developed. It is established that size-dependent quantization of vibrational modes results in formation of excess pressure of the phonon gas $P_{ph}$ acting outwards the crystal. Based on the concept of the phonon gas pressure, size dependence of thermodynamic properties of nanocrystals was described, and size influence on a shift of a phase transformation temperature was explained.

## 1. Introduction

Nanoparticles and nanofilms show strong size dependence of physical and thermodynamic properties. Size reduction of a crystal results in changes in both temperature and heat of phase transformations, shifts of the Curie point and the Neel point in the case of magnetic materials, increase in the temperature of the amorphous-to-crystalline transition, increase in diffusion rate and conductivity, a rise in the temperature of the transition to the superconducting state, acceleration of chemical and electrochemical processes on the particle surface and other effects.

For historical reasons, most efforts were concentrated on studies of size effects using melting of nanocrystals as an example. Reduction of the melting temperature of free nanoparticles was first observed in x-ray and electron-microscopy studies of granular films of Pb, Sn, Bi [1]; Pb, In [2]; Ag, Cu, [3], and Au [4]. Subsequent experimental and theoretical investigations of thermodynamic properties of nanocrystals (for reviews, see [5–8]) allowed to establish simple empirical relationships between these properties and particle size. Particularly, it was found that melting temperature $T_m$ of free nanoparticles in mesoscopic size range is inversely proportional to the particle size $l$,

$$T_m = T_{m,b}(1 - a/l), \qquad (1)$$

where $T_{m,b}$ is the bulk melting temperature, $a > 0$ is a dimensional factor depending on material and shape of the nanoparticle. A similar relation describes reduction of the melting heat with decreasing of nanocrystal size [9–12].

Subsequently, it was established that such thermodynamic properties as isobaric and isochoric heat capacities, Debye temperature etc., also vary with crystal size [13–14]. In addition to affecting the melting transition, reduction of particle size in some cases leads to formation of new polymorphic modifications of the crystal. Here, two situations are possible: in the first case, a polymorphic modification is formed at lower temperature than in bulk specimens (e.g. formation of the fcc phase in films and small particles of Co [8]), while, in the second case, crystal size reduction is accompanied by formation of phases that are not observed in bulk specimens (fcc phases in thin films of Nb, Ta, Mo and W [8]). New polymorphic modifications are usually observed in systems obtained by thin films deposition [8] or grinding of solids [15–16]. A similar size-induced effect takes place during amorphization of the matter in thin films (for reviews, see [7, 8]). The above examples, undoubtedly, attest to direct influence of the crystal size on the state of the matter inside the crystal. Due to small thickness of the surface layer in crystals [17, 18], this influence cannot be ascribed to surface effect on the bulk properties of a nanoparticle, i.e. to changes in the surface energy under a change of the phase state of the particle [5–8]. Recently, it was demonstrated [19,20] that size reduction of a crystal is accompanied by quantization (and increasing) of its vibrational energy, making the pressure of the phonon gas of the crystal increase. A force corresponding to such excess pressure of the phonon gas is directed outwards the crystal, i. e. additional pressure in a nanoparticle is negative. In the present study we show that it is the negative pressure of the phonon gas that is responsible for size-dependent effects in nanocrystals. Analytic expressions are obtained relating thermodynamic properties of a nanocrystal with its size.

## 2. Phase transitions in nanocrystals

To describe phase equilibrium in a nanocrystalline system, we proceed from the fact that, in simple systems (including nanocrystals of the mesoscopic size), the phase state is determined by two thermodynamic quantities: pressure $P$ and temperature $T$. Therefore, a change in the temperature of a phase transition in mesoscopic nanocrystals can result from a change of the pressure in the system. Let us determine pressure and temperature changes accompanying formation of a nanoparticle according to the principles of thermodynamic description [21]. When a nanocrystal is formed from a bulk crystal, an interface with surface tension $\sigma$ is created. This can result in arising additional pressure $P_\sigma$ in the particle. Besides, due to finite size $l$ of the nanoparticle, eigenvibrations with the wavelengths $\lambda > 2l$ are not excited, and the energies of vibration modes with wavelengths $\lambda < 2l$ depend on the crystal size. Such size-induced quantization of the energies of vibration modes $\varepsilon_{vibr}$ leads to presence of size-dependent pressure of the phonon gas $P_{ph} \sim -\partial f_{vibr}/\partial l$ in a nanoparticle [19]. The force associated with the phonon gas pressure $P_{ph}$ in free nanocrystals is directed outwards the crystal, resulting in a decrease of the total pressure in the particle. Therefore, formation of a nanocrystal is accompanied by a pressure change $\Delta P = P_\sigma + P_{ph}$, which consists of the surface pressure $P_\sigma$ $l$ and phonon gas pressure $P_{ph}$ $l$. A total pressure change $\Delta P$, caused by a nanoparticle formation from a bulk phase, can be either positive or negative, affecting substantially a temperature shift of phase transition in the nanocrystal.

On the phase transformation curve $T_m = T_m$ $P$ , the chemical potentials of two phases are equal,

$$\mu_s\left(T_m,P\right) = \mu_l\left(T_m,P\right). \qquad (2)$$

Differentiating (2) with respect to temperature with regard to the equation $T_m = T_m(P)$ leads to the Clausius-Clapeyron equation [21],

$$\frac{dT}{dP} = \frac{T_m(v_l - v_s)}{q_m}$$

which determines relationship between a change in the phase transition temperature $\Delta T = T_m - T_{m,b}$ and a change in the pressure $\Delta P$ in the system,

$$T_m = T_{m,b} + \left(\frac{T_{m,b}(v_l - v_s)}{q_{m,b}}\right)\Delta P \qquad (3)$$

where $q_{m,b} = T_{m,b}(s_l - s_s)$ is specific heat of the phase transformation for the bulk phase, $v_j$ and $s_j$ are molecular volume and entropy of a bulk phase, respectively ($j = s$ for solid phase and $j = l$ for a liquid). In fact, equation (3) represents the first term of the power series expansion of the phase transformation temperature with respect to $\Delta P$.

In the phase transformation curve of nanoparticles, it follows from the equality of chemical potentials of both phases

$$\mu_j = \varepsilon_j - Ts_j + P_j v_j + \frac{\sigma_j^* S_j}{\rho_j V_j},$$

that $q_m$ equals to the difference of specific enthalpies $w_j = \varepsilon_j + P_j v_j + \frac{\sigma_j^* S_j}{\rho_j V_j}$ of two phases [21], and thus $q_m$ is given by

$$q_m = q_m^0 + \Delta P_l v_l - \Delta P_s v_s + \frac{\sigma_l^*}{\rho_l}\left(\frac{S_l}{V_l}\right) - \frac{\sigma_s^*}{\rho_s}\left(\frac{S_s}{V_s}\right), \qquad (4)$$

where $\varepsilon_j$ is molecular energy of the $j$-th phase, $\Delta P_j = P_j - P_0$ is excessive pressure in the nanoparticle in the state $j$, $\sigma_j$ is surface tension at the interface, $S_j$ and $V_j$ are surface area and volume of the particle, respectively, $\rho_j V_j = N$ is the number of atoms (molecules) in the particle, $\rho_j$ is a density of the phase $j$. If we neglect the difference in $\varepsilon_j$ and $v_j$ for the bulk and nanocrystalline phases in (4), then $q_m^0 = \varepsilon_l - \varepsilon_s + P_0(v_l - v_s)$ is specific heat of the phase transformation in the bulk phase, $\sigma_j^* = \sigma_j - T\frac{d\sigma_j}{dT}$ is specific surface energy of the nanoparticle.

The obtained equation for $q_m$ explicitly expresses size dependence of the heat of the phase transformation of nanocrystals.

It is generally believed that a crystal melts when it reaches anharmonic instability state determined by changes in temperature and pressure [22]. When a system approaches the instability point, a series of its parameters (heat capacity, thermal expansion coefficient, rms displacement of atoms etc.) exhibits nonlinear rise. On the other hand, the formation energy of lattice structure defects decreases dramatically [23, 24], resulting in a sharp increase in the defect concentration and breakdown of the crystalline order being the main cause of melting, i.e. structural disordering of the crystal. Such picture of the physical nature of melting allows one to give credence to applicability of one-phase melting models, based on the mechanism of the anharmonic lattice instability: Lindemann criterion, Born criterion etc.

As follows from (3), the cause of a change of the melting temperature in the nanomatter lies in a pressure change $\Delta P$ in a nanoparticle which brings the anharmonic crystalline phase into unstable state. The crystal-liquid transition in the nanoparticle is accompanied by a change in the atomic (molecular) energy, a pressure jump between the solid and liquid phase, and a change in the surface tension at the solid-liquid interface. All these changes contribute to the melting heat of nanocrystals (4).

There is a problem in determination of surface pressure $P_\sigma\, l$ in a nanoparticle. In the case of a liquid droplet, when the liquid phase is in equilibrium with its vapour, $P_\sigma\, l$ is the Laplace pressure. For nanomatter being in equilibrium with its vapour (e.g., a liquid droplet), surface tension at the interface determines the particle shape and leads to formation of excessive Laplace capillary pressure [21]. However, in the case of crystals, the vapour pressure of the nanomatter is extremely small, and thermodynamic equilibrium between the matter and its vapour is not reached. As a result, surface relaxation of a nanocrystal takes place with a constant number of surface atoms because the surface energy depends on atomic density on curved surface of the crystal [17]. This relaxation leads to formation of a strained surface layer in a nanocrystal, compressing (or expanding) the crystal [17].

Therefore, a distinctive feature of nanocrystalline matter is presence of both excessive pressure $P_{ph}$ of the phonon gas and surface pressure $P_\sigma$, whose value depends on nanoparticle size and affects substantially mechanical and thermodynamic properties of the nanosystem.

## 3. Pressure of the phonon gas

Reduction of a crystal size $l$ is accompanied by rearrangement of its vibrational spectrum. For crystals of mesoscopic size, the long-wavelength part of the spectrum with wave vectors $k \sim \pi/l$ is mostly affected. In this range of vibrations that is described by wave equations of theory of elasticity, the spectrum is discrete with increment of $\Delta k \sim \pi/l$ increasing as the nanoparticle size decreases.

According to the laws of statistical mechanics [25], high-temperature distribution of coordinates of atomic displacements $q_{j,\alpha}$ ($j = 1, 2, ..., N; \alpha = x, y, z$) in a nanocrystal is given by [26, 27, 28]

$$f_N(q_1, q_2, ..., q_N) = e^{-\frac{\Delta U}{k_B T}} = \prod_{j=1,2,...,N} \exp\left[-\frac{\beta_1}{k_B T} \sum_{\alpha=x,y,z} n_{j,\alpha} q_{j,\alpha}^2\right]. \quad (5)$$

Here $\Delta U$ is a change in the potential energy of atoms due to their harmonic displacements from the equilibrium positions, $\beta_1$ is a parameter of quasi-elastic bond between neighbouring atoms, and coefficients $n_{j,\alpha}$ are determined with the vibrational spectrum,

$$n_{j,\alpha} = \frac{1}{2} \sum_{\mathbf{k}} \tilde{\omega}_{\mathbf{k}}^2 e_{j,\alpha}(\mathbf{k}) e_{j,\alpha}(\mathbf{k}), \quad (6)$$

where $e_{j,\alpha}(\mathbf{k})$ is $\alpha$-th projection of an eigenvector of the dynamical matrix, $\tilde{\omega}_{\mathbf{k}}$ is reduced frequency of the $\mathbf{k}$-th vibration mode,

$$\tilde{\omega}_{\mathbf{k}} = \sqrt{\frac{M}{\beta_1}} \omega_{\mathbf{k}}.$$

When computing coefficients $n_{j,\alpha}$ for nanocrystals in the mesoscopic size range [20, 28], we took into account the fact that vibrations with wavelengths $\lambda > 2l$ ($k_{min} \geq \pi/l$) can not be excited in such systems, and that the vibration spectrum has quasi-discrete character in the long-wave part. For atoms located in the bulk of a nanocrystal, distribution of displacements from the lattice sites is spherically symmetrical ($n_{j,x} = n_{j,y} = n_{j,z} = n(l)$), and it follows from equation (6) that [20]

$$n(l) = n_0 \left(1 + \gamma \frac{R}{l}\right), \quad (7)$$

where $n_0$ is the value of the parameter $n(l)$ for the bulk crystal (e.g., $n_0 = 2$ for fcc crystals [26]), $R$ is interatomic distance, $\gamma \sim 1$ is a parameter depending on material, shape and surrounding of the nanoparticle. It should be especially emphasized that expression (7) applies to the nanocrystal of mesoscopic size only. With decreasing particle size, the expression for the $n(l)$ dependence is more complex, and for nanoclusters consisting of several tens atoms $n(l) \to n_0$ [28]. To calculate the equilibrium free energy, phonon gas pressure and thermodynamic properties of nanocrystals, let us employ a variational approach [26, 27] developed to study thermodynamic properties of bulk crystals and extended also to the description of nanocrystals [19, 20, 28]. To write down the free energy functional we need approximated potential of interatomic interaction. It is convenient to choose the three-parameter Morse potential,

$$u(r) = A\left[e^{-2\alpha(r-R_0)} - 2e^{-\alpha(r-R_0)}\right]. \tag{8}$$

The three free parameters $A, \alpha, R_0$ to be determined from experimental data ensure good approximation of the shape of potential wells around the lattice sites [26]. This feature, together with simple form of the potential (8), allows one to use the Morse potential to describe not only thermodynamic properties of simple crystals at ambient pressure [23, 26], but also properties of crystals under high pressure [27]. At high temperature, a nanocrystal of mesoscopic size consisting of $N$ atoms can be described with the functional of the Helmholz free energy [19, 20]

$$f(\tau,v,c,b) = \frac{F}{AN} = \left\{(\frac{\tau}{3}+3\tau\log\frac{c\Lambda}{\tau}) + \frac{z}{2}\left[e^{-2b+\frac{2\tau}{n l c^2}} - 2e^{-b+\frac{\tau}{2n l c^2}}\right] - \frac{a_3\tau^2}{c^6}\left[e^{-2b+\frac{2\tau}{n l c^2}} - \frac{1}{4}e^{-b+\frac{\tau}{2n l c^2}}\right]^2\right\} + \frac{K}{(\alpha R_0+b)^2} + \left(\frac{\sigma S}{AN}\right) \tag{9}$$

where

$$K = \frac{3}{5}\left(\frac{3\pi^2}{2}\right)^{2/3}\frac{\hbar^2\alpha^2}{m_e k_B A}.$$

In equation (9), $\tau = k_B T / A$ is reduced temperature, $\Lambda = \hbar\alpha(AM)^{-1/2}$ is the de Boer parameter for the Morse potential, $S$ is surface area of the particle. The first term defines the entropy part of the free energy of vibrations, the second term is the average potential energy of interatomic interaction, and the third term is a contribution to the free energy due to the cubic anharmonicity of vibrations. In the case of simple metals, the expression (9) for the free energy should include an additional contribution due to kinetic energy of the electron gas,

$$\varepsilon_{el} = \frac{3}{5}\left(\frac{3\pi^2}{2}\right)^{2/3}\frac{\hbar^2}{m_e^*(\alpha R_0+b)^2 A}, \tag{10}$$

where $m_e^*$ is effective electron mass. The parameters for Au are given in Table 1 ($m_e$ is electron mass). The last term in (9) describes a contribution of the surface energy to the free energy of the nanocrystal. As mentioned above, a number of surface atoms remains unchanged during the process of attaining the equilibrium in nanocrystals, and relaxation of the surface layer leads to changes in distances between neighbouring atoms only. The value of this change varies with the surface curvature, affecting the surface energy $\sigma$, i.e. $\sigma$ should depend on the particle size. In the case of a flat surface, there is no size dependence.

**Table 1.** Parameters of the Morse potential (8) for Au [30] used in the calculations.

| $A/k_B$ (K) | $R_0$ (Å) | $\alpha$ (Å$^{-1}$) | $a_3$ | $m_{eff}/m_e$ |
|---|---|---|---|---|
| 6965 | 2.78 | 2.13 | 1.86 | 1.18 |

The reduced variational parameters $b = \alpha(R - R_0)$ and $c = \sqrt{\beta_1/A\alpha^2}$ define lattice expansion and quasi-elastic bond of atoms, respectively. Their equilibrium values are determined from the following conditions:

$$\left.\frac{\partial f}{\partial b}\right|_{\tau,c} = 0, \qquad (11)$$

$$\left.\frac{\partial f}{\partial c}\right|_{\tau,b} = 0. \qquad (12)$$

The temperature range where equation (9) is valid is determined by the inequality $\tau > c\Lambda$.

It should be noted that use of the variational approach [26, 27] simplifies substantially the procedure of calculation of the equilibrium properties of crystals, especially, for anharmonic crystals with complicated form of interatomic interaction or crystals exposed to external forces.

We use the free energy functional (9) to examine thermodynamic properties of nanocrystals in the mesoscopic size range assuming that variation of the parameters $b$ and $c$ in the surface layer can be neglected, and all the surface effects are taken into account with the parameter $\sigma$.

Let us consider nanocrystals with the fcc lattice with volume per atom being equal to $v = \left(R_0 + b/\alpha\right)^3/\sqrt{2}$. We will consider both spherical nanoparticles of radius $l$ and nanofilms of thickness $l$. In the case of nanofilms, the last term in (9) depends neither on the variational parameters nor on the film thickness. For spherical particles, this term is $\propto (\alpha R_0 + b)^3/l$, i. e. it depends both on the variational parameter and on the particle radius. In this case, $\sigma$ is not surface tension $\sigma_0$ at an interface between a bulk crystal and vacuum, but it is surface tension appearing owing to a change in interaction of atoms at curved surface of the nanoparticle [17]. It is obvious that $\sigma \ll \sigma_0$.

With the chosen potential of interatomic interaction (8), an approximate solution of the system of equations (11)–(12) can be found. As a result, the equilibrium values of the variational parameters $c_0$ and $b_0$ are given by:

$$c_0(\tau,l) \approx \left(\frac{\tau}{\tau_c(l)}\right)^{1/4} \left[\left(\frac{n\,l}{p}\right)\sin\left(\frac{\pi}{6} - \frac{\arctan\left(\frac{\tau_c}{\tau}-1\right)^{1/2}}{3}\right)\right]^{-1/2}, \qquad (13)$$

$$b_0(\tau,l) \approx \frac{3\tau}{2c_0^2 n(l)} + \left(\frac{K}{6(\alpha R_0)^3} - \frac{a_3\tau^2}{6c_0^6} - \frac{3R_0^2\sigma}{4\sqrt{2}Al\alpha}\right)\exp\left(\frac{\tau}{c_0^2 n(l)}\right), \quad (14)$$

where

$$p \approx 1.172 - \frac{2K}{3(\alpha R_0)^3} + \frac{3}{\sqrt{2}}\frac{R_0^2\sigma}{A}\frac{1}{\alpha l} + \frac{0.00032 a_3 n(l)^3}{\tau} \quad (15)$$

In figures. 1 and 2 we plotted, as an example, the equilibrium values of the variational parameters $c_0$ and $b_0$ versus crystal size at different temperatures. As seen from the figures, these parameters show maximal sensitivity to the particle size in the vicinity of the melting temperature of the particle.

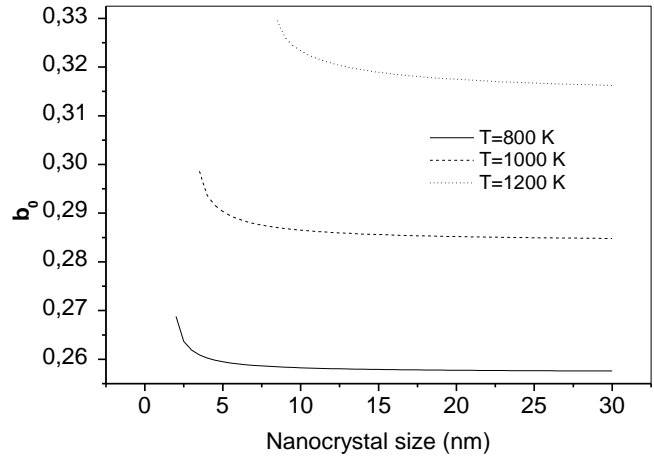

**Figure1.** Reduced interatomic distance $b_0(l)$ as a function of radius of a spherical gold nanoparticle.

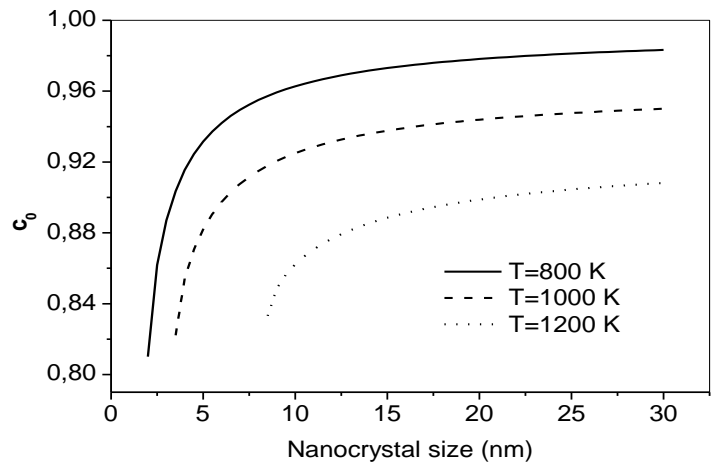

**Figure2.** Quasi-elastic bond parameter $c_0(l)$ as a function of radius of a spherical gold nanoparticle.

A crystal can be heated to critical temperature which is determined by simple expression,

$$\tau_c = \frac{4p^3}{qa_3 n_0^3 \left(1+\gamma \frac{R_0}{l}\right)^3}, \qquad (16)$$

where

$$q \approx 1.095 - \frac{0.91K}{(\alpha R_0)^3} + \frac{9}{2\sqrt{2}} \frac{R_0^2 \sigma}{A} \frac{1}{\alpha l} + \frac{0.00043 a_3 n(l)^3}{\tau}. \qquad (17)$$

As follows from (13), the value of the quasi-elastic bond parameter becomes imaginary at high temperature $\tau > \tau_c$, i. e. the vibrational subsystem of the crystal becomes unstable. It is necessary to point out that the instability temperature $\tau_c$ (15) is determined mainly by the cubic vibrational anharmonicity. In the vicinity of $\tau_c$, the quasi-elastic bond parameter is finite and behaves asymptotically as

$$c_0\big|_{\tau \to \tau_c} = \sqrt{\frac{2p}{n\,l}} \left(\frac{\tau}{\tau_c}\right)^{1/4} \left(1 - \frac{1}{\sqrt{3}}\sqrt{1-\frac{\tau}{\tau_c}}\right)^{-1/2}. \qquad (18)$$

A finite value of the quasi-elastic bond parameter is indicative of stability of the crystal lattice in the entire temperature range up to the melting point.

Based on numerical analysis of the functional (9), it was shown [19] that reduction of the crystal size is accompanied by an increase in the pressure $P_{ph}$ of the phonon gas, resulting in total pressure in a nanocrystal being negative. It was concluded that in [19] $P_{ph}$ is responsible for a shift of the phase transformation temperature in nanocrystals and size dependence of thermodynamic properties.

If the surface pressure $P_\sigma\, l$ is discarded, the difference between a nanocrystal and a bulk crystal is reduced to the presence of the excessive phonon gas pressure $P_{ph}$ that should be taken into account for a nanocrystal. In other words, thermodynamic properties of the medium in nanocrystals are equivalent to the properties of a bulk crystal ($l \to \infty$ or $\gamma \to 0$) exposed to external pressure $P_{ph}$. Therefore, nanocrystalline medium should be described by the Gibbs free energy $\Phi = F + P_{ph}V$, and the condition of minimum of $\Phi$ with respect to $b$ yields the following expression for $P_{ph}$:

$$P_{ph} = -\frac{A}{\partial v/\partial b} \left\{ \frac{\partial f}{\partial b}\bigg|_{b_0,c_0,\gamma=0} - \frac{\partial f}{\partial b}\bigg|_{b_0,c_0,\gamma \neq 0} \right\}, \qquad (19)$$

where $b_0$ and $c_0$ are the equilibrium values of internal parameters of a nanocrystal with $\gamma \neq 0$, and the second term in (19) equals to zero in view of (11). Expanding the second term in brackets into a series in $R/l \ll 1$, that is equivalent to expanding into a series in $\gamma$, and retaining only the first term linear $\gamma$, we obtain

$$P_{ph} = -4\sqrt{2}\gamma \frac{R_0}{l}\alpha \frac{k_B T}{(R_0+b_0/\alpha)^2}\varphi_1\left(c_0,b_0\right), \qquad (20)$$

where

$$\varphi_1\left(c_0,b_0\right) = \frac{1}{c_0^2}\left\{-\frac{1}{4}e^{-b_0+\frac{\tau}{4c_0^2}} + e^{-2b_0+\frac{\tau}{c_0^2}} - \frac{a_3\tau^2}{192c_0^6}\left(e^{-2b_0+\frac{\tau}{2c_0^2}} - 30e^{-3b_0+\frac{5\tau}{4c_0^2}} + 128e^{-4b_0+\frac{2\tau}{c_0^2}}\right)\right\}, \qquad (21)$$

is a function depending weakly on size and temperature of the nanoparticle (figure 3). In figures 4 and 5 we present size and temperature dependences of the phonon gas pressure in a nanocrystal. While the size dependence of $P_{ph}$ is rather well described by the hyperbolic function $l$, the temperature dependence of $P_{ph}$ shows a number of specific features. At low temperatures, the linear rise of the pressure with temperature suggests that the phonon gas is ideal. As temperature approaches the melting point, the vibration anharmonicity results in interaction between phonons and essentially nonlinear behavior of $P_{ph}$.

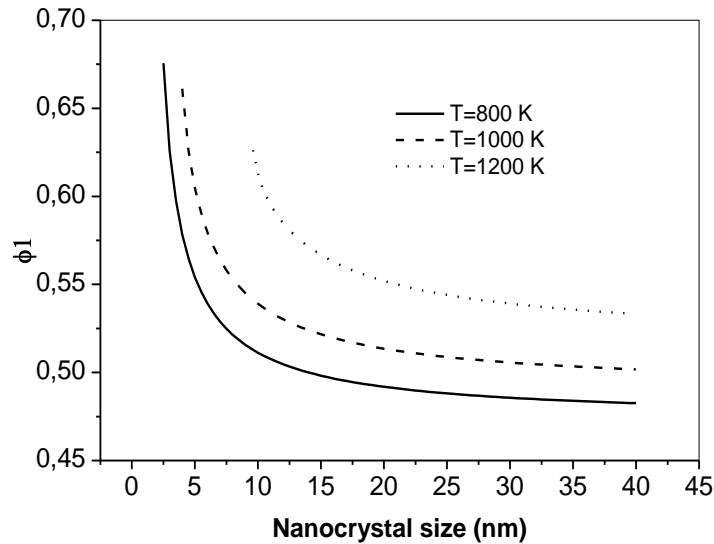

**Figure 3.** The parameter $\varphi_1$ as a function of size for Au nanocrystals.

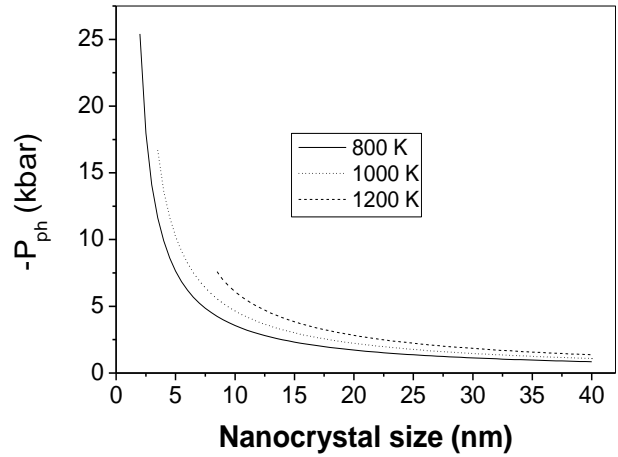

**Figure 4.** Size dependence of the phonon gas pressure $-P_{ph}$ in Au nanocrystals calculated from (20) at different temperatures.

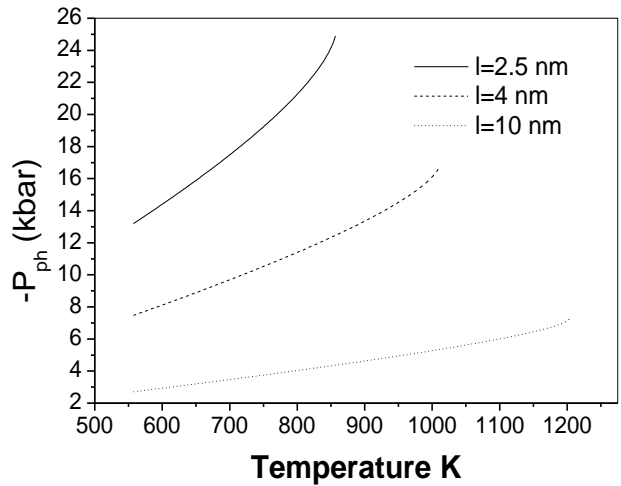

**Figure 5.** Temperature dependence of the phonon gas pressure $-P_{ph}$, calculated from (20) at different nanocrystal sizes.

## 4. Melting of nanocrystals

Assuming that excessive pressure in nanocrystals is related to the phonon gas pressure only (20), and assuming $P_\sigma = 0$, and using equation (3) which relates a change in the phase transformation temperature and a change in the pressure in the system, we can obtain an explicit expression for size dependence of the phase transformation temperature:

$$T_m = \frac{T_{m,b}}{\left(1 + 4\gamma \frac{R_0}{l} \alpha R_0 \frac{k_B T_{m,b}}{q_m} \frac{v_l - v_S}{v_S} \varphi_1 \right)} \quad . \tag{22}$$

Under assumptions adopted, equation (22) is strict in the range $l \gg R_0$, with $q_m = q_{m,b}$. However, as follows from (4) and experimental results (see, e.g., [9–12]) $q_m$ is also size-dependent. According to (4), for spherical particles, this dependence is written by

$$q_m = q_{m,b}\left[1 - \frac{\Delta P_S - \Delta P_l}{q_{m,b}}v_S - \frac{3(\sigma_S^* - \sigma_l^*)}{q_{m,b}\rho_S}\frac{1}{l}\right]. \qquad (23)$$

In equation (23), it is assumed that $v_S \approx v_l$ and $\rho_S \approx \rho_l$. Since $\sigma_S^* > \sigma_l^*$ for most systems, then the last term in (23) contributes to reduction of $q_m$ as the particle size decreases. While the excessive pressure $\Delta P_S$ in a crystalline nanoparticle is due to the phonon gas (20), calculation of the pressure $\Delta P_l$ in a liquid particle is rather complicated. However, all contributions to $\Delta P_j$ are of the order of $1/l$, and corrections due to $\Delta P_j$ can be determined from experimental data.

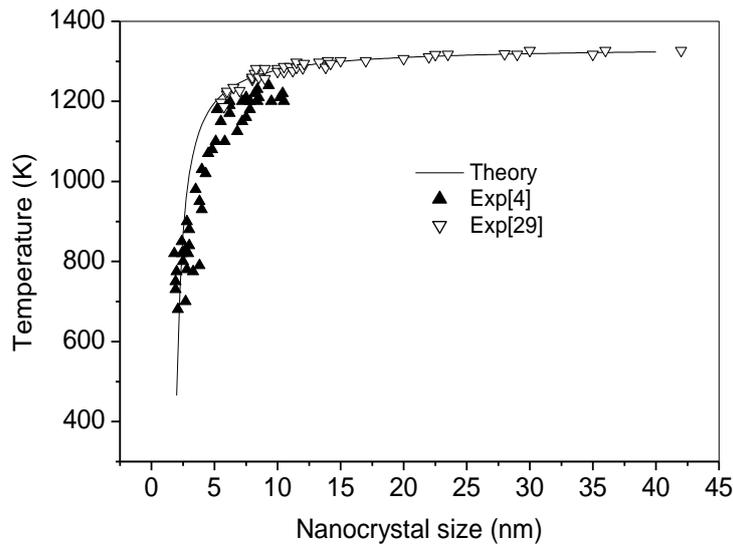

**Figure 6.** Size dependence of the melting temperature of free spherical particles of gold. The curve was obtained from (22) with respect to $q_m$ $l$ dependence.

In figure 6, we show the size dependence of the melting temperature of gold nanoparticles calculated from (22) with respect to $q_m$ $l$ dependence. The following physical parameters of gold were used [4]:

**Table 2.** Parameters used for calculation of the melting curve of Au nanocrystals.

| $q_{m,b}$ (kJ mole$^{-1}$) | $\sigma_S$ (J m$^{-2}$) | $\dfrac{d\sigma_S}{dT}$ (J m$^{-2}$ K$^{-1}$) | $\sigma_l$ (J m$^{-2}$) | $\dfrac{d\sigma_l}{dT}$ (J m$^{-2}$ K$^{-1}$) | $\rho_S$ (kg m$^{-3}$) | $\rho_l$ (kg m$^{-3}$) | $\Delta v/v_S$ |
|---|---|---|---|---|---|---|---|
| 12.41 | 1.38 | $-4.33\cdot 10^{-4}$ | 1.13 | $-1.0\cdot 10^{-4}$ | $1.84\cdot 10^{4}$ | $1.728\ 10^{4}$ | $5.1\ 10^{-2}$ |

For these values of parameters, assuming $\Delta P_S = \Delta P_l$, we find

$$q_m = q_{m,b}\left(1 - \frac{\beta}{l}\right), \tag{24}$$

where $\beta \approx 1.8$ nm. It should be noted that taking the $q_m(l)$ dependence for gold nanocrystals contributes remarkably to $T_m$ only for $l \leq 10$ nm. The second term in the denominator in (22) is much less than unity, so (22) can be represented in the form of (1) with

$$a = 4\gamma\alpha R_0^2 \frac{k_B T_{m,b}}{q_m} \frac{v_l - v_S}{v_S}\varphi_1. \tag{25}$$

## 5. Interatomic distance in crystals

Numerous x-ray and electron-microscopy studies of nanocrystals ( [31–35]) suggest that size reduction of a nanocrystal is accompanied by relatively small (about 0.1% in order) change of the interatomic distance. Depending on surrounding conditions and surface state of a specimen, both lattice expansion and contraction are observed. In many cases, embedding a specimen into a matrix or deposition of foreign atoms on its surface results in a crossover from the size-related expansion of the lattice to contraction. As follows from (14), there are two mechanisms that may make lattice contract or expand as the crystal size decreases. These are enhancement of the anharmonicity due to softening of vibration modes of the nanocrystal (reduction of $c_0(l)$) and contraction/expansion of the lattice due to stress in the surface layer. While the vibration anharmonicity is determined by the interatomic potential and lattice geometry, the surface layer stress depends both on surface condition and presence of foreign atoms adsorbed at the surface. For illustrative purposes, in figure 7 we present theoretical size dependence of reduced interatomic distance calculated at different coefficients of the surface tension. In the absence of a stressed layer ($\sigma = 0$), increasing interatomic distance is due to the phonon gas pressure $P_{ph}$ making a nanocrystal expand. Enhancement of the surface tension leads to "capillary" compression of the nanocrystal ($\sigma_S = 0.72$ J m$^{-2}$), though the capillary pressure is much less than the phonon pressure and has little effect on the shift of the phase transformation temperature in the nanocrystal. Besides, an intermediate case of non-monotonic dependence of $b(l)$ is possible ($\sigma_S = 0.3$ J m$^{-2}$), when increase in $b(l)$ near the melting temperature of the nanocrystal is due to additional nonlinear increase in pressure of the phonon gas (figure 5).

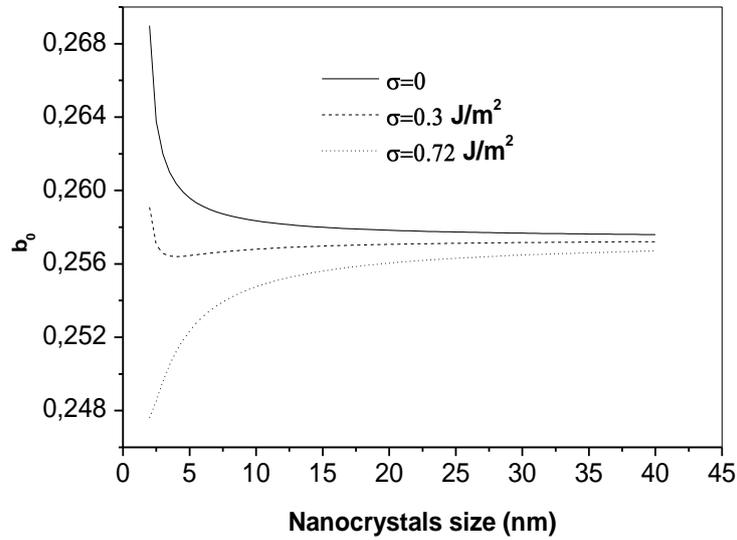

**Figure 7.** Size dependence of reduced interatomic distance $b = \alpha(R - R_0)$ in Au nanocrystals.

### 6. Equation of state of nanocrystals

Additional pressure of the phonon gas in a nanocrystalline medium affects equation of state of nanocrystals determining a relationship between pressure applied to the medium and its volume. In the case of bulk crystals, one of such equations is the Mie-Grüneisen equation [36, 37]

$$P_0 = -\left(\frac{\partial F}{\partial V}\right)_T = -\frac{\partial E_0}{\partial V} + \frac{\gamma_G E_T}{V}, \qquad (26)$$

where $E_0$ is the energy of zero vibrations, $E_T$ is the energy of thermal vibrations, $\gamma_G$ is the Grüneisen parameter. Since basic thermodynamic properties depend on crystal size, a corresponding contribution to $P_0$ should appear for a nanocrystalline medium. It should be noted that this contribution is not directly related to $P_{ph}$, which changes internal parameters of the medium and is partially relaxed.

To determine pressure in the bulk of a nanoparticle, let us introduce a new parameter $x = \dfrac{\tau}{c^2 n \, l}$ instead of $c$. Then the functional (9) for the free energy functional can be rewritten as

$$f\left(\tau,v,x,b\right) = \frac{1}{3}\tau - \frac{3}{2}\tau\log\left(x\right) + 3\tau\log\left(\frac{\Lambda}{\sqrt{n_0\tau}\sqrt{1+\gamma R_0/l}}\right) + 6\left(e^{-2b+2x} - 2e^{\left(-b+\frac{x}{2}\right)}\right) -$$
$$\frac{a_3 x^3 n_0^3 \left(1+\gamma R_0/l\right)^3}{\tau}\left(e^{-2b+2x} - \frac{1}{4}e^{\left(-b+\frac{x}{2}\right)}\right)^2 + \frac{K}{\left(\alpha R_0 + b\right)^2} + \frac{3\sigma}{\sqrt{2}A\alpha^3 l}\left(\alpha R_0 + b\right)^3 \quad (27)$$

and the variational condition (12) is replaced with

$$\left(\frac{\partial f}{\partial x}\right)_{\tau,b} = 0. \quad (28)$$

The pressure $P$ acting in the medium whose state is described with the free energy functional (27) is given by

$$P_n = -\left(\frac{dF}{dV}\right)_\tau = -\frac{AN}{dV/dl}\left\{\left(\frac{\partial f}{\partial x}\right)_{\tau,x_0,b_0}\frac{\partial x}{\partial l} + \left(\frac{\partial f}{\partial b}\right)_{\tau,x_0,b_0}\frac{\partial b}{\partial l} + \left(\frac{\partial f}{\partial l}\right)_{\tau,x_0,b_0}\right\} \quad (29)$$

Two first terms in (29) are equal to zero due to (11) and (28). From (29), we obtain

$$P_n = -\frac{3N}{2\,dV/dl}\frac{1}{l^2}\left\{\frac{R_0\gamma k_B T}{\left(1+\gamma\frac{R_0}{l}\right)}\left(1 + \frac{2a_3 n\left(l\right)^3 x^3\left(-\frac{1}{4}e^{-b+\frac{x}{2}} + e^{-2b+2x}\right)^2}{\tau^2}\right) - \frac{2\sigma R_0^3}{\sqrt{2}}\right\}. \quad (30)$$

A factor before the parenthesis in (30) depends on shape of the nanoparticle. In the case of plate-shaped nanoparticles, $V = Sl/2$ ($S = \text{const}$), $N = \rho V$, $\rho = 1/v$, where $v$ is volume per atom,

$$P_n = -\frac{3}{2}\gamma\rho\frac{R_0}{l}\frac{k_B T}{\left(1+\gamma\frac{R_0}{l}\right)}\left(1 + \frac{2a_3 n\left(l\right)^3 x^3}{\tau^2}\left(-\frac{1}{4}e^{-b+\frac{x}{2}} + e^{-2b+2x}\right)^2\right). \quad (31)$$

For spherical nanoparticles of radius $l$, with the assumption $\sigma = 0$, we get

$$P_n = -\frac{1}{2}\gamma\rho\frac{R_0}{l}\frac{k_B T}{\left(1+\gamma\frac{R_0}{l}\right)}\left(1 + \frac{2a_3 n\left(l\right)^3 x^3}{\tau^2}\left(-\frac{1}{4}e^{-b+\frac{x}{2}} + e^{-2b+2x}\right)^2\right). \quad (32)$$

In figure 8 we plotted pressure $-P_n$ as a function of radius of a gold nanoparticle calculated from (32) for different temperatures.

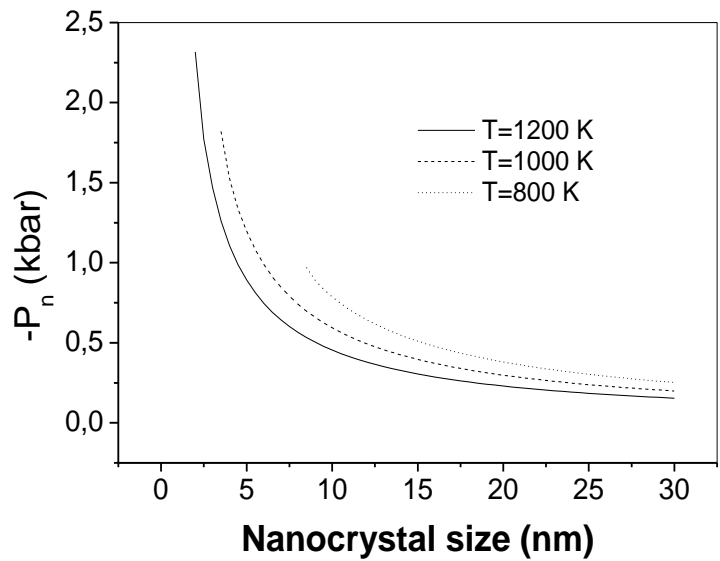

**Figure 8.** A correction to the equation of state (26) versus radius of a gold nanoparticle.

In view of $P_n$, equation (26) takes on the form

$$P_0 = -P_n - \frac{\partial E_0}{\partial V} + \frac{\gamma_G E_T}{V}. \tag{33}$$

Hence it follows, in particular, that crystal volume increases at $\sigma = 0$. Recently, an attempt was made to expand the Mie-Grüneisen equation to nanocrystalline systems [38]. Determination of $P_n$ was based on separation of contributions of surface and volume parts of vibration spertum to the free energy of a nanocrystal.

### 7. Conclusion

A hypothesis about presence of negative pressure in thin films was first formulated in Ref. [39] for possible explanation of size dependence of thermodynamic and physical properties of thin films. However, absence of proper theoretical and experimental background suspended development of work in this direction for a long time. Preference was given to the "capillary" model which related a change in the phase transformation temperature with the jump of the surface energy in phase transitions in nanocrystals (review is given in [5–8]). However, apparent simplicity of the "capillary" model disguised, for a long time, a fundamental mistake. Indeed, a first order phase transition in a thermodynamic system occurs when the system reaches a point of instability (due to temperature or pressure change). In the case of melting, the instability of crystalline state is due to anharmonicity of collective atomic vibrations [22–24], so a shift of the phase transition temperature (3) is related to a pressure change in the crystalline phase only. Jumps of specific volume, molecular energy, surface energy of a nanoparticle and pressure during the transition from one phase to another contribute to the melting specific heat (4) only, which is determined from the condition of equality of chemical potentials in the equilibrium curve of nanophases. As a consequence, size influence on the phase transition (melting) temperature in a nanocrystal is related to both changes in the excess pressure $\Delta P$ in the nanocrystal and size dependence of the heat of melting $q_m(l)$.

Reduction of crystal size to mesoscopic or nanoscopic size range is accompanied by changes in the mechanical, thermodynamic, electric and structure properties of the crystals. As crystal size decreases, a number of twins increases in the crystal, thus improving mechanical properties of the crystal [40]. In nanocrystals, a number of vacancies is increased [41], processes of amorphization of the medium are facilitated [42], as well as restoring of nanogranulated material after plastic deformation [43–44]. As was already indicated, both size and surrounding medium also affect substantially thermodynamic properties and phase transition temperature of a nanosized medium. Apparently, all these effects point to the fact that size reduction triggers physical processes that affect the state of the medium in nanoparticles. Since thermodynamic state of a one-component medium is determined by its pressure and temperature only, then it is a rise in the phonon gas pressure that governs the change of thermodynamic properties of a nanocrystal when its size decreases. In this context, it is hard to agree with the claims forming the basis of "capillary" models of nanocrystal melting [1–9]. A change in the surface energy at phase transition in a nanocrystal affects the heat of transition only (4) and can affect the temperature of phase transition, particularly, melting, in an indirect way only (19).

**Acknowledgement**

This work was supported in part by Award no 16/12-H in the framework of the Complex Program of Fundamental Investigations "Nanosized system, nanomaterials, nanotechnology" of the National Academy of Sciences of Ukraine.